\documentclass[12pt]{iopart}
\usepackage{iopams}
\usepackage{graphicx}

\begin{document}

\title[Thermal expansion and magnetostriction of RAl$_3$]{Thermal expansion, heat capacity and magnetostriction of RAl$_3$ (R = Tm, Yb, Lu) single crystals.}

\author{S L Bud'ko$^1$, J C Frederick$^1$, E D Mun$^1$, P C  Canfield$^1$ and G M  Schmiedeshoff$^2$}

\address{$^1$Ames Laboratory US DOE and Department of Physics and Astronomy, Iowa State University, Ames, IA 50011, USA}

\address{$^2$Department of Physics, Occidental College, Los Angeles, CA 90041, USA}

\begin{abstract}
We present thermal expansion and longitudinal magnetostriction data for cubic RAl$_3$ (R = Tm, Yb, Lu) single
crystals. The thermal expansion coefficient for YbAl$_3$ is consistent with an intermediate valence of the Yb ion,
whereas the data for TmAl$_3$ show crystal electric field contributions and have strong magnetic field
dependencies. de Haas-van Alphen-like oscillations were observed in the magnetostriction data of YbAl$_3$ and LuAl$_3$,
several new extreme orbits were measured and their effective masses were estimated. Zero and 140 kOe specific heat
data taken on both LuAl$_3$ and TmAl$_3$ for $T \leq 200$ K allow for the determination of a CEF splitting scheme
for TmAl$_3$.
\end{abstract}

\pacs{65.40.-b, 75.20.Hr}
\submitto{\JPCM}
\maketitle

\section{Introduction}

Rare earth trialuminides, RAl$_3$, have been studied for several decades. The crystal structure of these materials
is very sensitive to the rare earth ionic radius and it changes from hexagonal for light rare earths to cubic
(Cu$_3$Au - type) for TmAl$_3$, YbAl$_3$, LuAl$_3$ (and ScAl$_3$) \cite{vuc65a},  with ErAl$_3$ reported to have
been synthesized in both crystallographic variants \cite{vuc65a,bar71a,bus68a}. TmAl$_3$ was reported to have a
singlet ground state \cite{bus68a,deu89a}: the magnetic susceptibility follows the Curie-Weiss law with the magnetic
moment corresponding to that of Tm$^{3+}$ at high temperatures and becomes a temperature-independent Van Vleck
type at low temperatures. YbAl$_3$, an intermediate valence compound \cite{hav73a} with the high Kondo
temperature, $T_K$, of 600-700 K, recently experienced a revival of interest to its physical properties
\cite{hie00a,ebi03a,bau04a,chr06a} based, in part, on the uncovering of a second, low temperature, energy scale,
Fermi liquid coherence ($T_{coh} \approx 40$ K), and slow crossover between $T_K$ and $T_{coh}$ \cite{cor02a}.

Whereas many physical properties were measured and analyzed for cubic trialuminides, thermal expansion and
magnetostriction data for these materials, in particular at low temperatures, appear to be absent from the
literature (except for an early report \cite{ian72a} on the thermal expansion of LuAl$_3$ and YbAl$_3$ between $\sim
90$ K and $\sim 800$ K). In addition, for TmAl$_3$ the exact crystal electric field (CEF) splitting of the Hund's
rule, ground state multiplet has remained unresolved \cite{deu89a,buc74a,wij70a,sug01a}, in part due to a lack of
high temperature and high magnetic field specific heat data. In this work we report extensive measurements of specific heat as well as thermal expansion and
magnetostriction data for RAl$_3$ (R = Tm, Yb, Lu) single crystals so as to compare with other salient physical
properties and to inquire into the effects of the CEF and intermediate valence (IV) on these characteristics.
Simple crystal structure and the ability to synthesize large, high quality, singly crystals facilitate this
endeavor.
\\
\section{Experimental methods}

\begin{figure}[tbp]
\begin{center}
\includegraphics[angle=0,width=120mm]{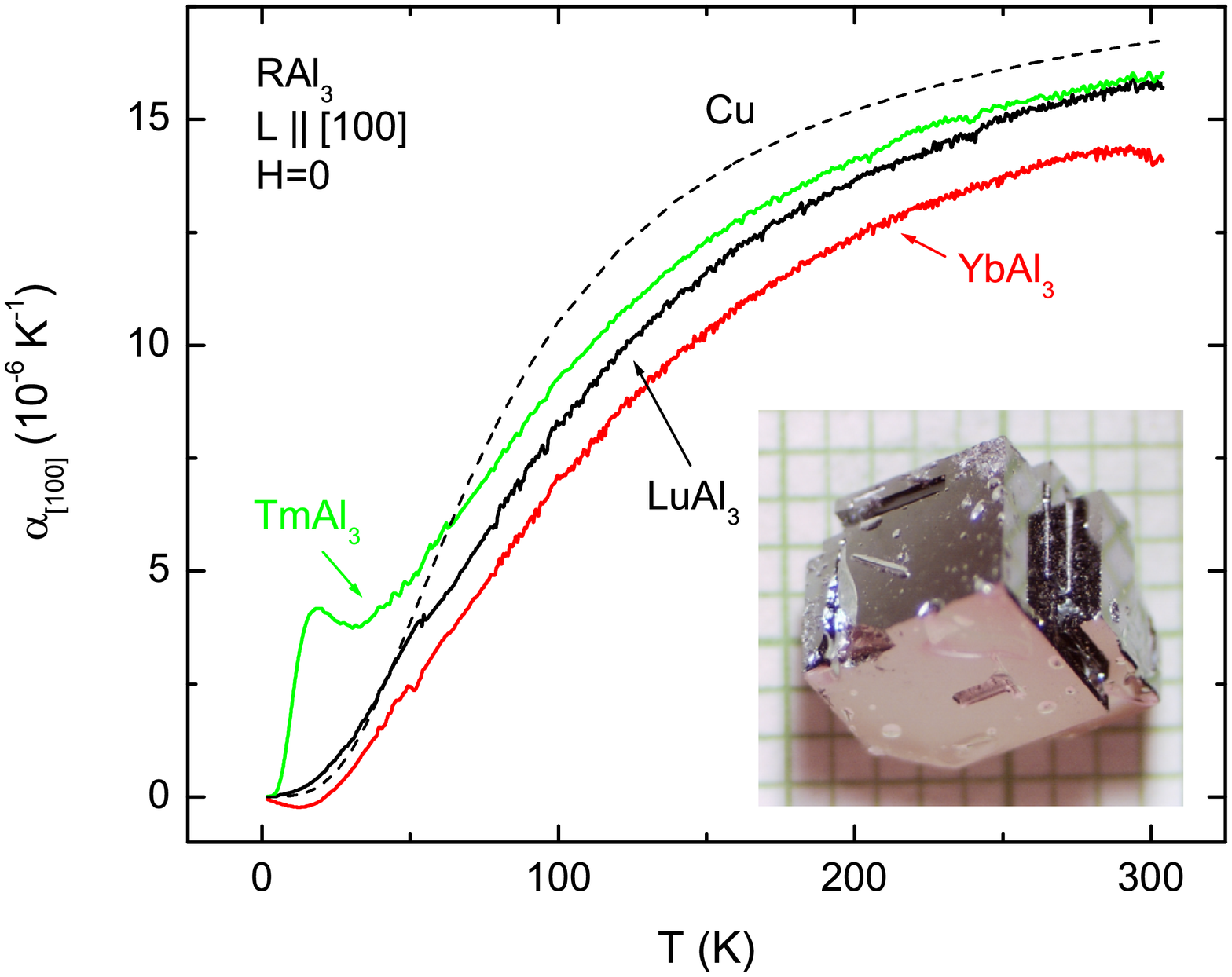}
\end{center}
\caption{(Color online) Temperature-dependent linear thermal expansion coefficients along $[100]$ direction of
RAl$_3$ (R = Tm, Yb, Lu) in zero applied field. Data for polycrystalline Cu \cite{kro77a} are shown as dashed line
for comparison. Inset: LuAl$_3$ crystal over a mm scale.} \label{F1}
\end{figure}

Large ($> 0.1$ cm$^3$) single crystals (Fig. \ref{F1}, inset) of RAl$_3$ (R = Tm, Yb, Lu) and R'$_{0.1}$Lu$_{0.9}$Al$_3$ (R' = Er, Tm,
Yb)  were grown from aluminum-rich binary (R - Al) or ternary (R' - Lu - Al) melts using a
self-flux method \cite{can92a}. High purity R (Ames Lab) and Al were placed, in atomic ratios of
R$_{0.12}$Al$_{0.88}$ or (Lu$_{0.9}$R$_{0.1}$)$_{0.12}$Al$_{0.88}$, in alumina crucibles.  These were sealed in
fused quartz ampoules under a 1/3 atmosphere partial pressure of Ar.  The ampoules were heated to $1050^\circ$C
and then cooled to $675^\circ$C over 150 - 160 hours, at which point they were removed from the furnace and the
excess Al was decanted. For R'$_{0.1}$Lu$_{0.9}$Al$_3$ with R' = Er, Tm the nominal concentration of R' was
corroborated by the Curie-Weiss fit of the high temperature susceptibility. The inset to Fig. \ref{F1} shows
crystals with clear, cubic morphology and mirrored $(100)$ facets. Linear dimensions as large as 7 mm were
commonly achieved.

Thermal expansion and magnetostriction were measured using a capacitive dilatometer constructed of OFHC copper; a
detailed description of the dilatometer is presented elsewhere \cite{sch06a}. The dilatometer was mounted in a
Quantum Design PPMS-14 instrument and was operated over a temperature range of 1.8 to 300 K in an applied magnetic
field up to 140 kOe. The samples were mounted in such a way that thermal expansion was measured along the $[100]$
direction. The applied magnetic field was also along $[100]$ such that $H \| L \| [100]$. The crystals were cut
and polished so that the typical distance between the parallel $(100)$ surfaces of the samples was $L_{[100]}
\approx 2-3$ mm. Heat capacity of the samples was measured using a hybrid adiabatic relaxation technique of the
heat capacity option in a Quantum Design PPMS-14.
\\

Thermodynamic properties of materials are frequently analyzed using the concept of a Gr\"{u}neisen function (or a
Gr\"{u}neisen parameter) \cite{bar99a}. For a single salient energy scale, $\varepsilon$, the Gr\"{u}neisen
parameter, $\gamma$, is defined as $\gamma = -d \ln \varepsilon/d \ln V$, where $V$ is a molar volume. Using
thermodynamic relations, we can obtain $\gamma(T,V) = \beta V/\chi_S C_p$, where $\beta$ is a volume thermal
expansion coefficient ($\beta = (\partial ln~V/\partial T)_P$), $\chi_S$ is an adiabatic compressibility ($\chi_S
= - (\partial ln~V/\partial P)_S$) and $C_p$ is a heat capacity at a constant pressure. For cubic materials
$\gamma(T,V) = 3 \alpha V/\chi_S C_p$, where $\alpha$ is a linear thermal expansion coefficient ($\alpha =
(\partial ln~a/\partial T)_P$, $a$ - is a lattice parameter). Sometimes, in the analysis of experimental data, 
lacking the temperature-dependent compressibility data, the temperature dependence of the Gr\"{u}neisen parameter
can be approximated \cite{pot81a} as being proportional to $\beta/C_p$ (or $3 \alpha/C_p$ for cubic materials)
under the assumption that the relative temperature dependence of $\chi_S$ is significantly smaller then that of
thermal expansion coefficient or heat capacity. We will adopt such approach in this work.

If more than one contribution to the thermodynamic properties is present (e.g. vibrational, electronic, magnetic,
etc.), the Gr\"{u}neisen parameters are not additive, rather the Gr\"{u}neisen parameter for the material is an
average, weighted by the heat capacity contribution of each component \cite{bar99a}: $\gamma = \sum\limits_{r}
\gamma_r C_r/\sum\limits_{r} C_r$.
\\

\section{Results and discussion}
\subsection{Thermal expansion and heat capacity}

Zero field, temperature-dependent, linear ($L \| [100]$) thermal expansion coefficients for RAl$_3$ (R = Tm, Yb,
Lu) (together with the literature data \cite{kro77a} of polycrystalline Cu for comparison), are shown in Fig.
\ref{F1}. In the overlapping temperature regions ($T > 90$ K), the thermal expansion coefficient values for
LuAl$_3$ and YbAl$_3$ deduced from the the lattice parameters as a function of temperature data \cite{ian72a} are
comparable to our results. At room temperature, thermal expansion values of the non-hybridizing TmAl$_3$ and
LuAl$_3$ are very similar. The differences in $\alpha(T)$ between these two materials, on cooling, and in
particular a peak in $\alpha(T)$ of TmAl$_3$ at $\sim 20$ K, are probably related to the CEF contributions to the
thermal expansion of TmAl$_3$. $\alpha(T)$ of YbAl$_3$ is lower than that of its non-magnetic analogue, LuAl$_3$,
over the whole temperature range. Such behavior is consistent \cite{woh81a} with YbAl$_3$ being an Yb-based,
intermediate valence material with a high, $T_K \gg 300$ K, Kondo temperature. Qualitatively, such behavior can be
understood by noting that the fractional Yb valence in YbAl$_3$ increases with increase in temperature (in the
temperature range of our measurements) \cite{sal75a} and that the ionic radius of Yb$^{3+}$ is smaller than that
of Yb$^{2+}$.
\\

\begin{figure}[tbp]
\begin{center}
\includegraphics[angle=0,width=120mm]{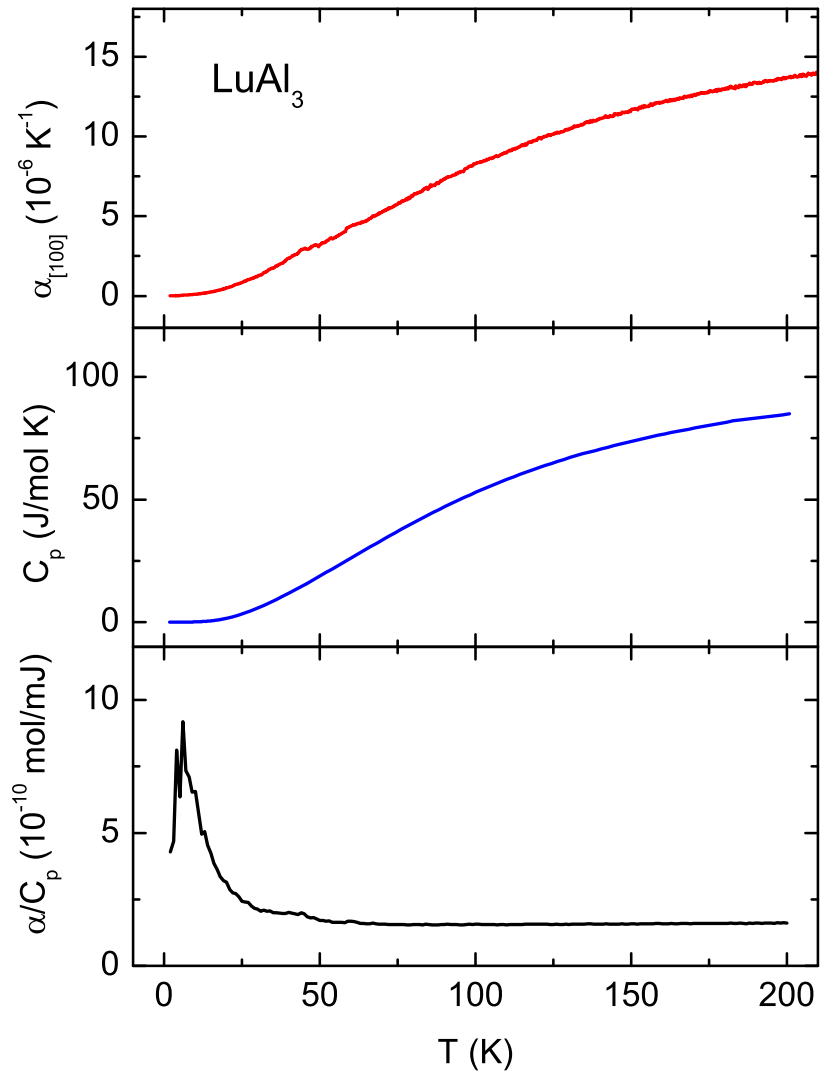}
\end{center}
\caption{(Color online) Temperature-dependent linear thermal expansion coefficient, heat capacity and the ratio
$\alpha/C_p$  for LuAl$_3$ in zero applied field.} \label{F2}
\end{figure}

Figure \ref{F2} presents the temperature-dependent linear thermal expansion coefficient ($\alpha$), the heat
capacity ($C_p$) and the ratio $\alpha/C_p$ for LuAl$_3$. $\alpha_{[100]}(T)$ and $C_p(T)$ have similar
temperature dependencies. The ratio of these two quantities (that, as mentioned above, is likely to approximate
the temperature dependence of the Gr\"{u}neisen parameter) is practically constant down to $\sim 50$ K, only
rising at lower temperatures and manifesting a peak at $\sim 8$ K. It is not unusual to observe a
temperature-dependent Gr\"{u}neisen parameter, even in simple, non-magnetic, metals \cite{bar99a}. This is due to
the different temperature dependencies of the electron and phonon contributions to the thermodynamic properties,
with these contributions becoming comparable in magnitude at low temperatures. Additionally, the error bars in
$\alpha/C_p$ may be somewhat enhanced at low temperatures as a result of calculating the ratio of two,
diminishingly small numbers.
\\

\begin{figure}[tbp]
\begin{center}
\includegraphics[angle=0,width=120mm]{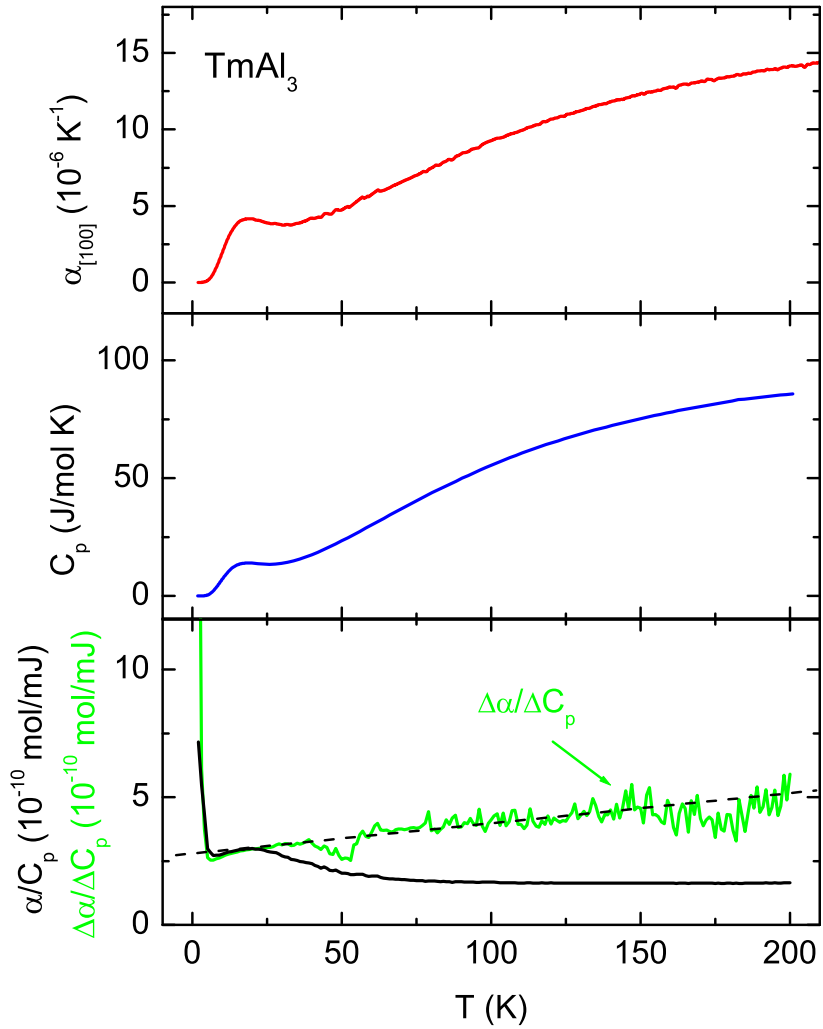}
\end{center}
\caption{(Color online) Temperature-dependent linear thermal expansion coefficient, heat capacity and the ratio
$\alpha/C_p$  for TmAl$_3$ in zero applied field. Additionally, on the lower panel $\Delta \alpha/\Delta C_p =
(\alpha($TmAl$_3) - \alpha($LuAl$_3))/(C_p($TmAl$_3) - C_p($LuAl$_3))$ is plotted. Dashed line is a guide for the
eye.} \label{F3}
\end{figure}

For TmAl$_3$ (Fig. \ref{F3}) both thermal expansion and heat capacity exhibit a broad peak in the $15 - 20$ K
temperature range, apparently related to CEF effects. This peak is still observable in the $\alpha/C_p$ curve,
however, if the data are plotted as $\Delta \alpha/\Delta C_p$, (i.e. after subtraction of the non-magnetic
background) it is not seen, confirming the same origin of the feature in $C_p$ and $\alpha$. The $\Delta
\alpha/\Delta C_p$ shows a close to linear, slowly changing with temperature, behavior, probably representing
rather well the behavior of the magnetic contribution to the Gr\"{u}neisen parameter.

\begin{figure}[tbp]
\begin{center}
\includegraphics[angle=0,width=120mm]{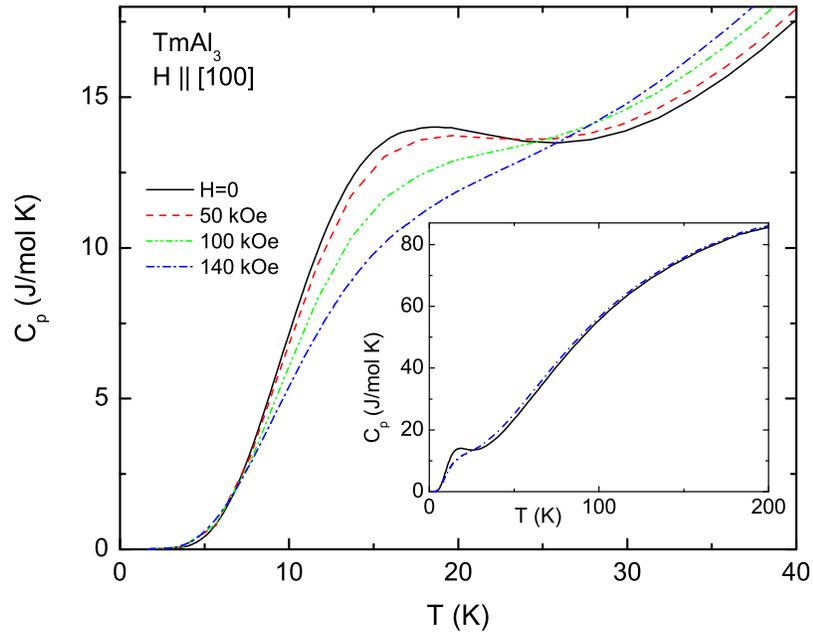}
\end{center}
\caption{(Color online) The low temperature part of heat capacity for TmAl$_3$ in applied fields $ H = 0, 50, 100,
140$ kOe. Field was applied along $[100]$ axis. Inset: data for $H = 0$ and 140 kOe up to 200 K.} \label{F3a}
\end{figure}

\begin{figure}[tbp]
\begin{center}
\includegraphics[angle=0,width=120mm]{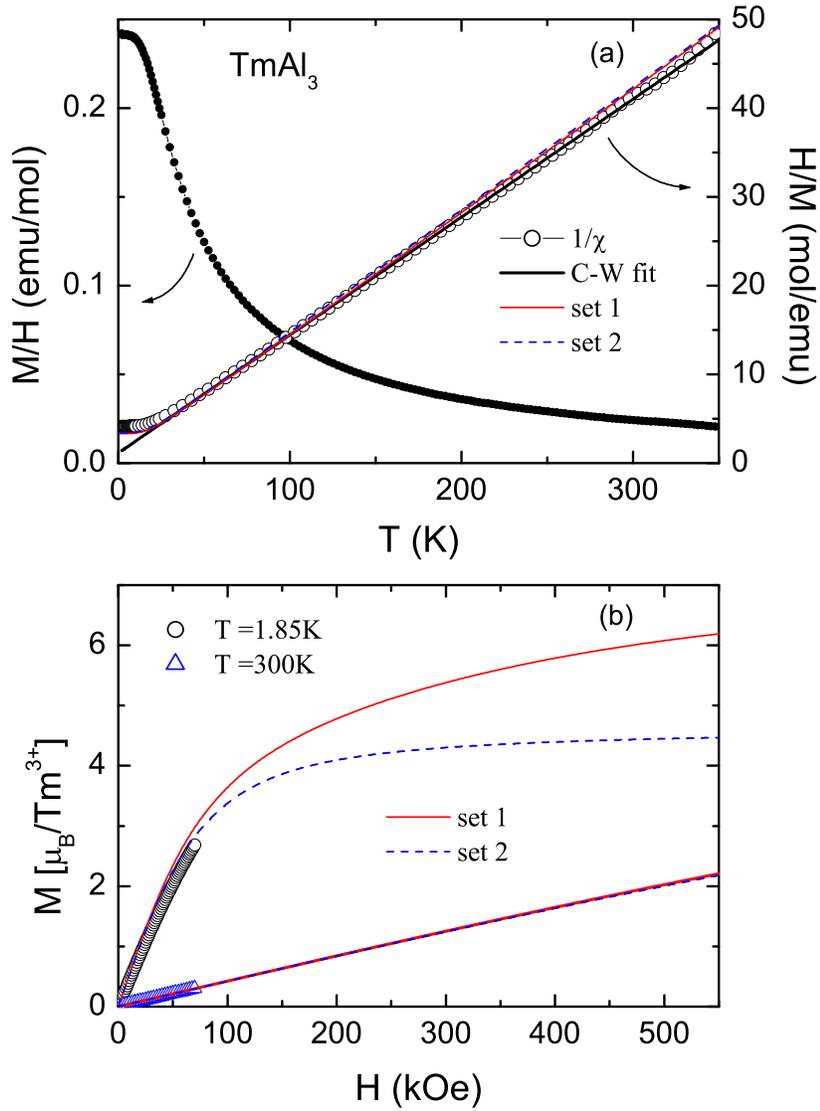}
\end{center}
\caption{(Color online) (a) Temperature-dependent magnetic susceptibility, $M/H$, inverse magnetic susceptibility,
$H/M$, taken at $H = 1$ kOe, $H \| [100]$ and (b) magnetization isotherms, $M(H)$, for $T = 1.85$ K and 300 K for
TmAl$_3$. Curie - Weiss fit of inverse magnetic susceptibility and CEF simulations of $H/M$ and $M(H)$ for two
possible sets of parameters (set 1: $W = -1.011, x = -0.298$ set 2: $W = -2.125, x = -0.682$) are show as lines.}
\label{F3aa}
\end{figure}

\begin{figure}[tbp]
\begin{center}
\includegraphics[angle=0,width=120mm]{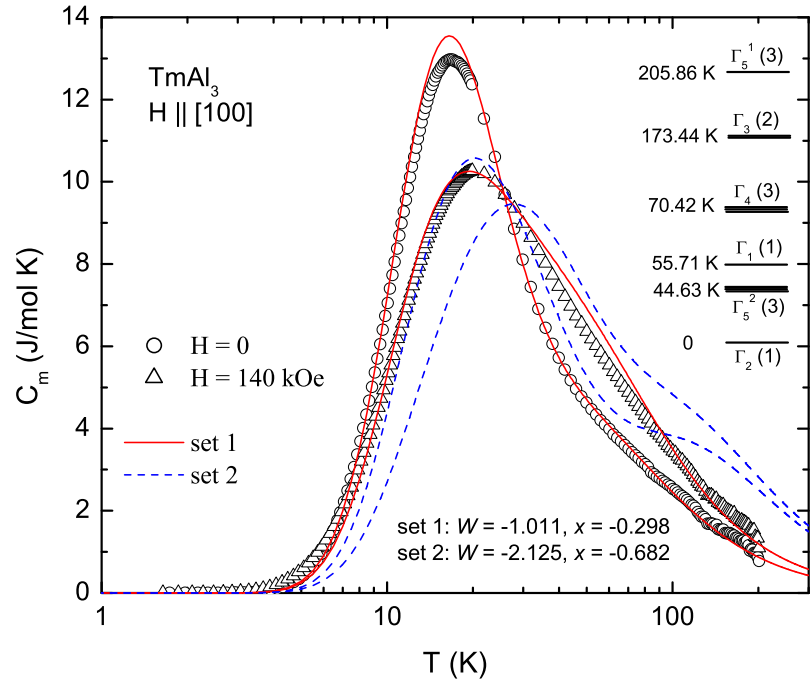}
\end{center}
\caption{(Color online) Magnetic contribution to the heat capacity of TmAl$_3$, $C_m = C_p^{TmAl_3} -
C_p^{LuAl_3}$, in zero and 140 kOe ($H \| [100]$) applied field. CEF simulations for two possible sets of
parameters are show as lines. Corresponding $W$ and $x$ values are listed for both sets are listed. CEF scheme for
set 1 is illustrated in the right side of the plot.} \label{F3b}
\end{figure}

The peak in heat capacity evolves with applied field (Fig. \ref{F3a}) reflecting the lifting of the degeneracy of
TmAl$_3$ energy levels. A variety of differing CEF schemes, based on fits of different experimental quantities
($C_p(T)$, or $\chi(T)$, or high field $M(H)$) were suggested for TmAl$_3$ in the literature
\cite{deu89a,buc74a,wij70a,sug01a}. To analyze our magnetization, susceptibility and specific heat data we will
use the approach delineated in Ref. \cite{lea62a}.  Since Tm$^{3+}$ ions in TmAl$_3$ have octahedron type of
coordination, both parameters, $W$ and $x$, of the CEF hamiltonian for cubic symmetry (see Ref. \cite{lea62a} for
definitions of these parameters and detailed discussion) are negative and the ground state is $\Gamma_2$ (set 1)
or $\Gamma_1$ (set 2). Temperature dependent susceptibility and magnetization isotherms up to 70 kOe (Fig.
\ref{F3aa}) allow for similar quality fits for CEF schemes with either ground states. Heat capacity data are
better fit with the CEF scheme corresponding to $\Gamma_2$ ground state (Fig. \ref{F3b}) (for both sets the $W$
and $x$ values from fits of temperature-dependent susceptibility and magnetic isotherms were used). The $W$ and
$x$ values for both sets are listed on the graph. An applied magnetic field changes the CEF splitting via the
Zeeman term. The $H = 0$ CEF levels plus this Zeeman term describe $H = 140$ kOe data well (Fig. \ref{F3b}).

\begin{figure}[tbp]
\begin{center}
\includegraphics[angle=0,width=120mm]{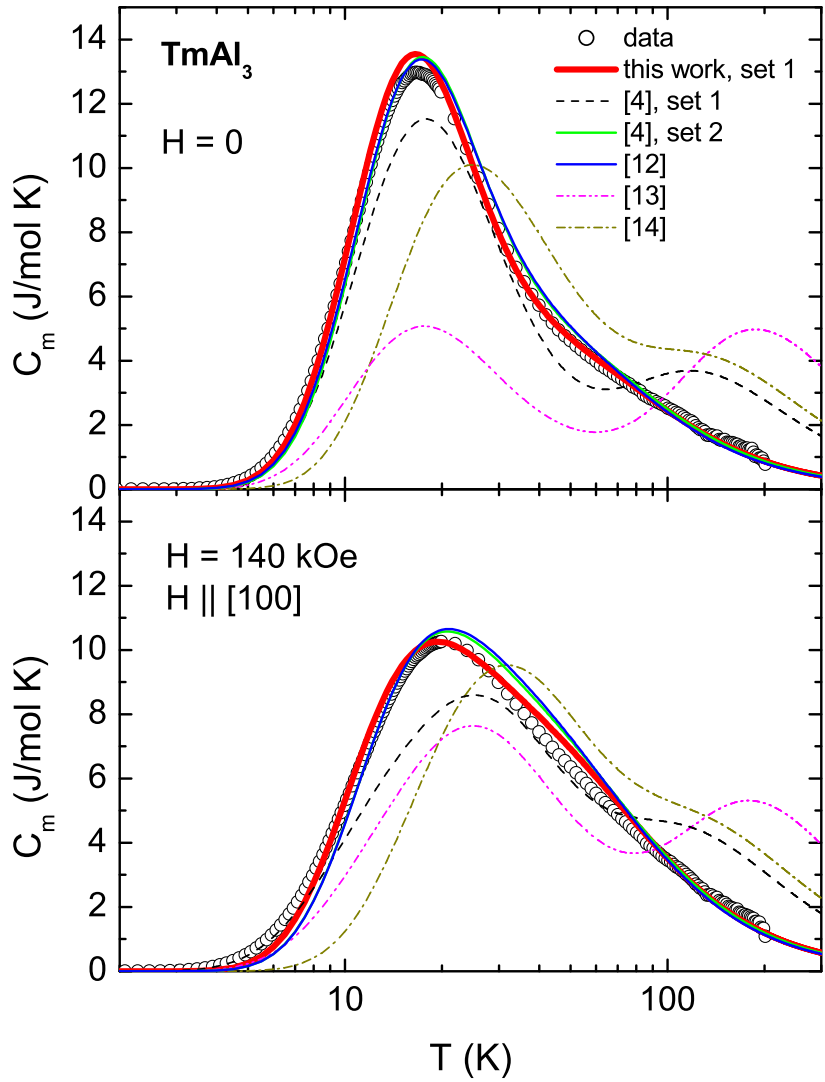}
\end{center}
\caption{(Color online) Magnetic contribution to the heat capacity of TmAl$_3$, $C_m = C_p^{TmAl_3} -
C_p^{LuAl_3}$, in zero and 140 kOe ($H \| [100]$) applied field. CEF simulations using set 1 from this work and
the literature data are show as lines. The following $W$ and $x$  values (listed as $\{W, x\}$) were used: this
work, set 1: $\{-1.011,-0.298\}$; Ref. \cite{deu89a}, set 1: $\{1.95, 0.82\}$, set 2: $\{-0.993, -0.282\}$; Ref.
\cite{buc74a}: $\{-0.94, -0.27\}$; Ref \cite{wij70a}: $\{3.24, -0.8\}$; Ref \cite{sug01a}: $\{-2.316, -0.827\}$}
\label{F3c}
\end{figure}

Fig. \ref{F3c} shows the simulation of the heat capacity behavior using the various CEF splitting schemes proposed
in the literature \cite{deu89a,buc74a,wij70a,sug01a} including two of the three schemes \cite{deu89a} presented as
indistinguishable given their low temperature data. Set 2 in Ref. \cite{deu89a} and the parameters from Ref.
\cite{buc74a} are close to our results (set 1) and describe the heat capacity data reasonably well, whereas the
parameters from Refs. \cite{wij70a,sug01a} and set 1 in Ref. \cite{deu89a} fail as reasonable approximations of
the experimental data.
\\

\begin{figure}[tbp]
\begin{center}
\includegraphics[angle=0,width=120mm]{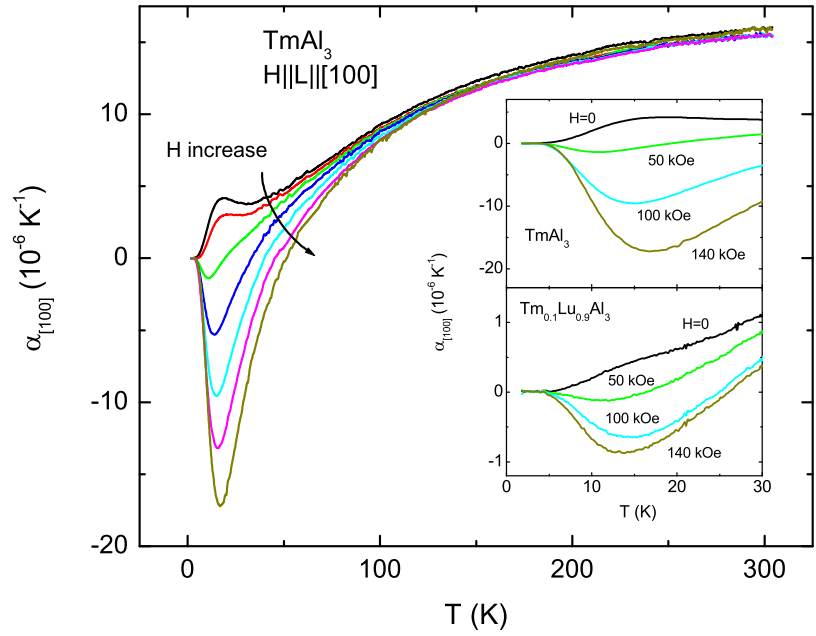}
\end{center}
\caption{(Color online) Temperature-dependent linear thermal expansion coefficient for TmAl$_3$ in applied fields
of $0, 25, 50, 75, 100, 125, 140$ kOe. Insets: low temperature part of $\alpha(T)$ in $0, 50, 100, 140$ kOe
applied fields for TmAl$_3$ and Tm$_{0.1}$Lu$_{0.9}$Al$_3$.} \label{F4}
\end{figure}

An applied field changes the behavior of the thermal expansion coefficient of TmAl$_3$ dramatically (Fig.  \ref{F4}).
In between 25 kOe and 50 kOe the low temperature maximum turns into a minimum. This minimum deepens in higher
fields and reaches $\approx - 1.7 \cdot 10^{-5}$ K$^{-1}$ near 15 K for $H = 140$ kOe. Clearly, as a result of the changing CEF
energy scales, the Gr\"{u}neisen parameter will be significantly different in the applied field.  Qualitatively
similar, but approximately an order of magnitude lower in size, field-induced changes in $\alpha_{[100]}(T)$ are
observed in Tm$_{0.1}$Lu$_{0.9}$Al$_3$ (cf. two insets to Fig. \ref{F4}). These perceptible field dependences of
$\alpha_{[100]}(T)$ in both TmAl$_3$ and Tm$_{0.1}$Lu$_{0.9}$Al$_3$ are consistent with a  CEF-related, single-ion
effect.

\begin{figure}[tbp]
\begin{center}
\includegraphics[angle=0,width=120mm]{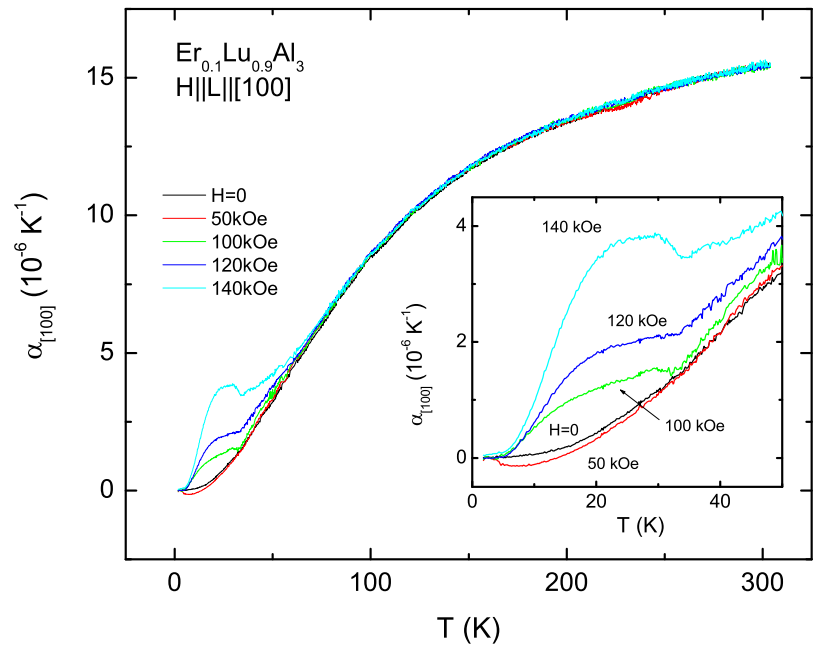}
\end{center}
\caption{(Color online) Temperature-dependent linear thermal expansion coefficient for Er$_{0.1}$Lu$_{0.9}$Al$_3$
in applied fields of $0, 50, 100, 120, 140$ kOe. Insets: low temperature part of $\alpha(T)$.} \label{F5}
\end{figure}

To illustrate the complexity of the effect of magnetic field on the thermal expansion coefficient,  data for
Er$_{0.1}$Lu$_{0.9}$Al$_3$ is shown in Fig. \ref{F5}. As opposed to Tm$_{0.1}$Lu$_{0.9}$Al$_3$ (Fig. \ref{F4},
inset) a {\it positive}, broad feature in $\alpha_{[100]}(T)$ below $\sim 40$ K grows with increasing applied
magnetic field. Transverse thermal expansion measurements in an applied field and some knowledge the CEF scheme
for Er$_{0.1}$Lu$_{0.9}$Al$_3$ would be needed for comparative analysis of the effects of magnetic field on these
two different materials.
\\

\begin{figure}[tbp]
\begin{center}
\includegraphics[angle=0,width=120mm]{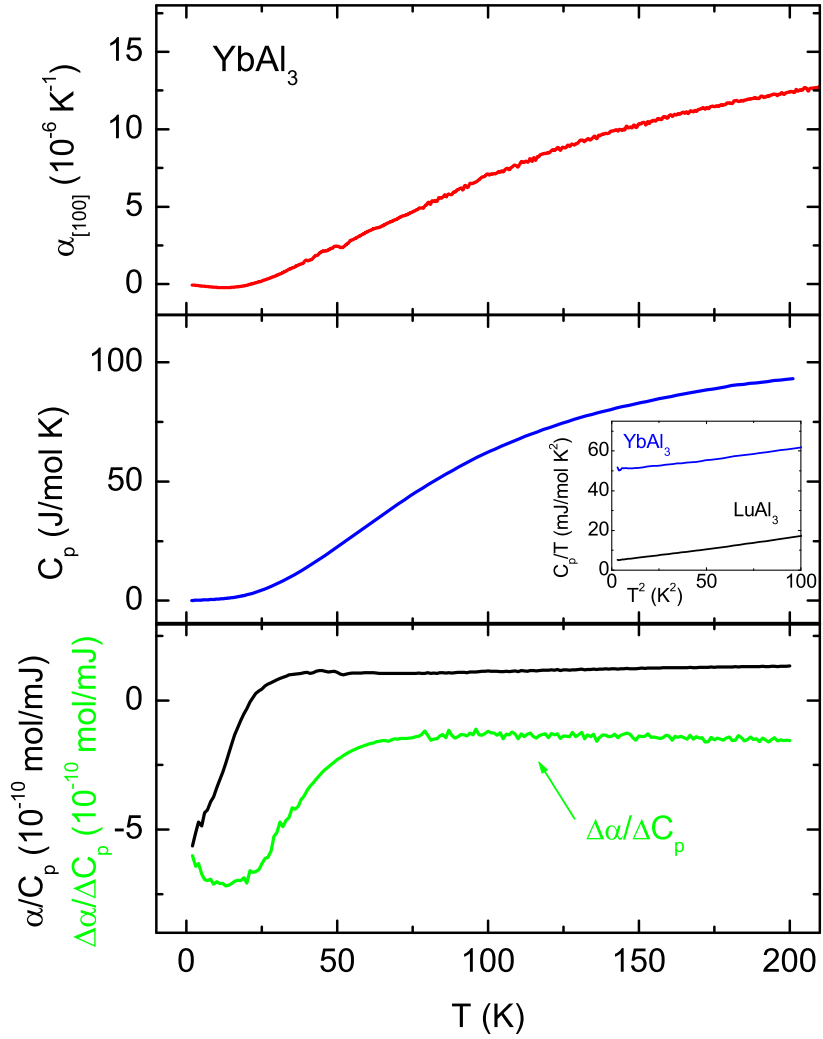}
\end{center}
\caption{(Color online) Temperature-dependent linear thermal expansion coefficient, heat capacity and the ratio
$\alpha/C_p$  for YbAl$_3$ in zero applied field. Additionally, on the lower panel $\Delta \alpha/\Delta C_p =
(\alpha($YbAl$_3) - \alpha($LuAl$_3))/(C_p($YbAl$_3) - C_p($LuAl$_3))$ is plotted. Inset: low temperature $C_p/T$
vs. $T^2$ for LuAl$_3$ and YbAl$_3$.} \label{F6}
\end{figure}

For YbAl$_3$, $\alpha_{[100]}(T)$ and $C_p(T)$ (Fig. \ref{F6}) have similar temperature dependencies, without any
apparent striking features. The ratio of these two quantities is practically constant down to $\sim 30$ K and then
decreases at lower temperatures. $\Delta \alpha/\Delta C_p$ is linear (and close to constant) at higher
temperatures and then, on further cooling, decreases and passes through broad minimum. Given that the Kondo
temperature is above room temperature \cite{hav73a}, it is tempting to try to connect these changes in  $\Delta
\alpha/\Delta C_p$ to the emerging second, low temperature ($\sim 30 - 40$ K), scale in YbAl$_3$ \cite{cor02a},
however at this point there is no clear evidence for such conjecture and more studies are required. Additionally, we cannot exclude this low temperature behavior to be the result of an interplay between phonons and (enhanced) electronic degrees of freedom.  Specific heat
data show an elevated electronic specific heat coefficient, $\gamma \sim 50$ mJ/mol K$^2$, for YbAl$_3$,
significantly higher than that for LuAl$_3$, in consistence with the intermediate valence nature of YbAl$_3$ and
similar to recently reported data \cite{urb07a}.

\begin{figure}[tbp]
\begin{center}
\includegraphics[angle=0,width=120mm]{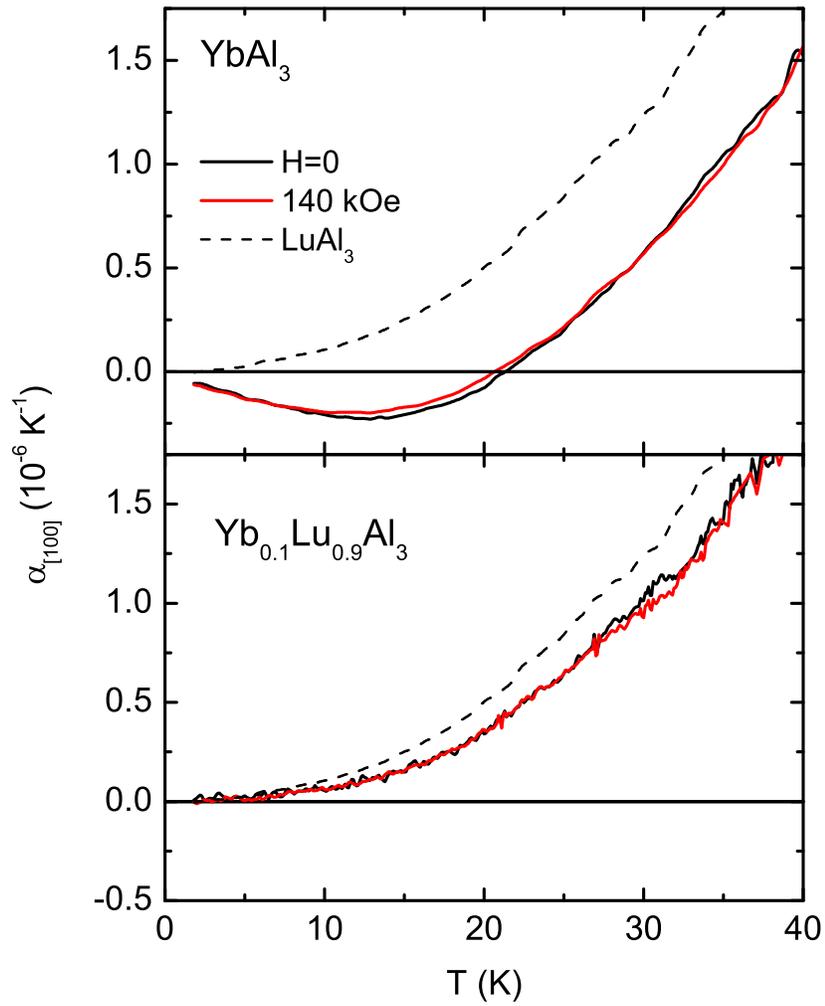}
\end{center}
\caption{(Color online) Low temperature part of linear thermal expansion coefficient for YbAl$_3$ and
Yb$_{0.1}$Lu$_{0.9}$Al$_3$ in applied fields of 0 and 140 kOe. For comparison, data for LuAl$_3$ are shown as
dashed line.} \label{F7}
\end{figure}

$\alpha_{[100]}(T)$ for YbAl$_3$ is not significantly affected by an applied magnetic field of 140 kOe (Fig.
\ref{F7}). This is not surprising for a material with a high, $600 - 700$ K, Kondo temperature. At low temperatures
($T \leq 20$ K) YbAl$_3$ exhibits a region of negative thermal expansion, $\alpha_{[100]}(T) < 0$. For
Yb$_{0.1}$Lu$_{0.9}$Al$_3$ the region of negative thermal expansion does not appear (above 1.8 K) since
$\alpha_{[100]}(T)$ is apparently dominated by the contribution from the LuAl$_3$ matrix. Both, in pure YbAl$_3$
and diluted Yb$_{0.1}$Lu$_{0.9}$Al$_3$, the temperature dependent thermal expansion is below the one for the
non-magnetic analogue, LuAl$_3$, consistent with an intermediate valence character of the Yb ion in pure and
diluted material.

\subsection{Magnetostriction and quantum oscillations}

\begin{figure}[tbp]
\begin{center}
\includegraphics[angle=0,width=120mm]{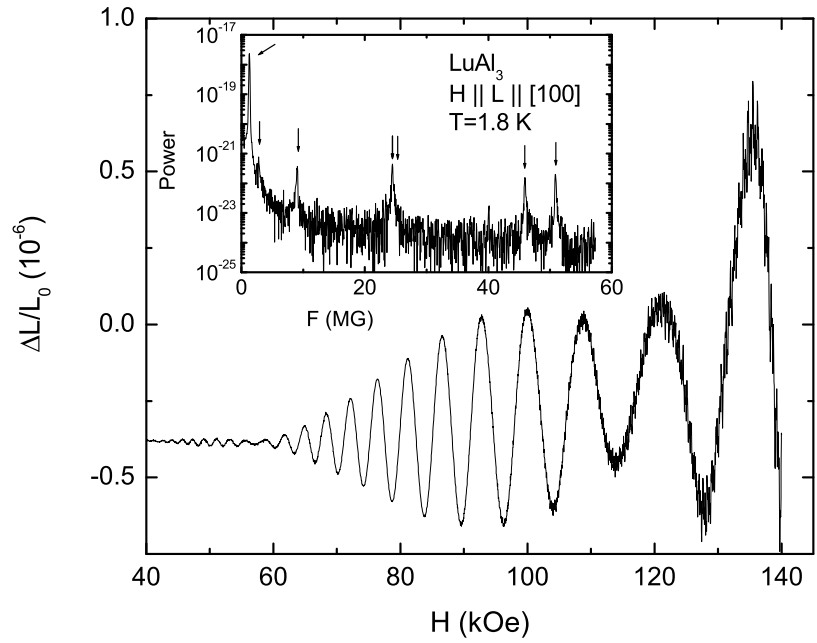}
\end{center}
\caption{High field part of longitudinal ($H \| L \| [100]$) magnetostriction in LuAl$_3$ measured at 1.8 K.
Inset: fast Fourier transform of the corresponding $\Delta L/L_0$ vs. $1/H$ data. Arrows mark the de Haas-van
Alphen frequencies.} \label{F8}
\end{figure}

\begin{figure}[tbp]
\begin{center}
\includegraphics[angle=0,width=120mm]{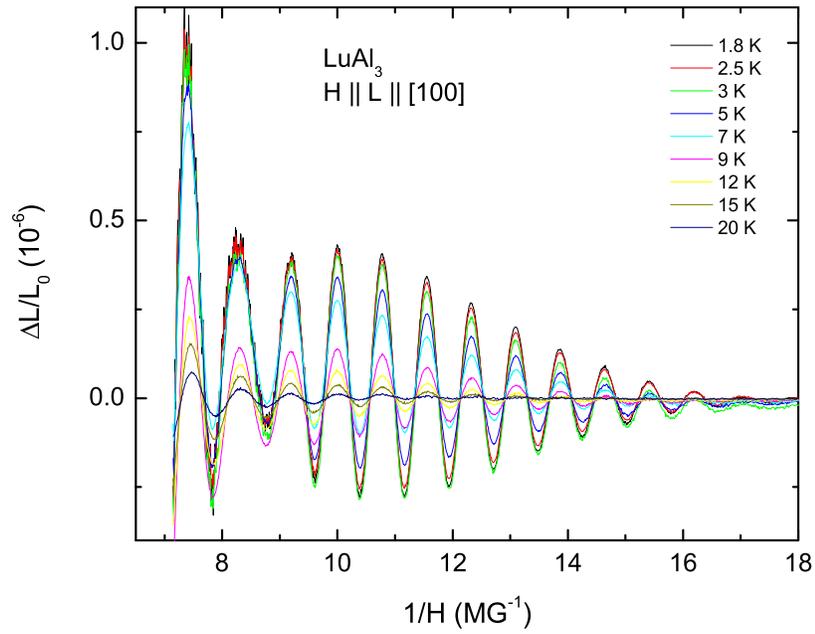}
\end{center}
\caption{(Color online) Longitudinal magnetostriction in LuAl$_3$ for $H \| [100]$ at different temperatures, from
1.8 K to 20 K plotted as a function of $1/H$. Constant background was subtracted from the data.} \label{F9}
\end{figure}

\begin{figure}[tbp]
\begin{center}
\includegraphics[angle=0,width=120mm]{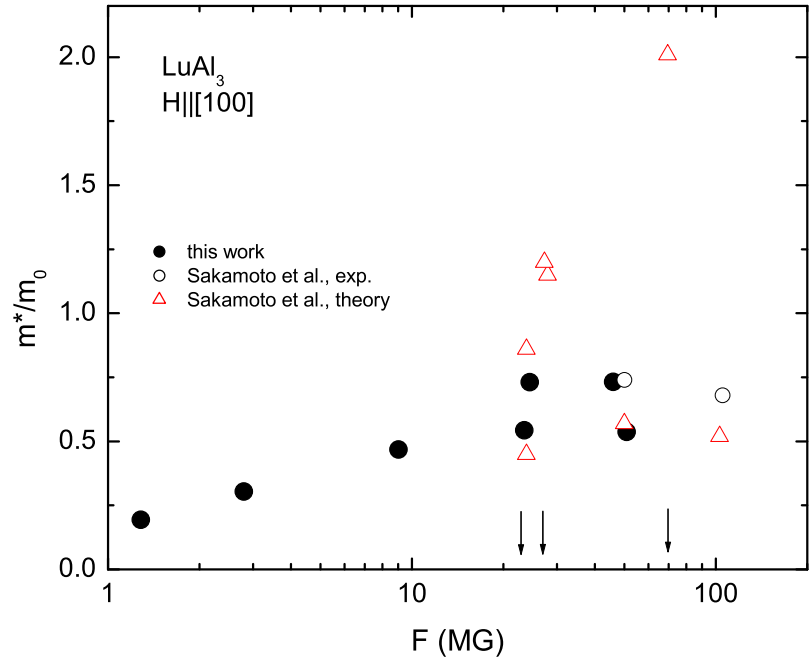}
\end{center}
\caption{(Color online) dHvA frequencies and corresponding effective masses for LuAl$_3$, $H \| [100]$. Filled
circles - this work, other symbols - literature data \cite{sak01a}: open circles - experiment, triangles - theory.
Note that Sakamoto et al. experimentally identified several other dHvA frequencies but did not determine the
corresponding effective masses. These experimental  frequencies are shown by arrows.} \label{F10}
\end{figure}

In LuAl$_3$ the base temperature (1.8 K) magnetostriction is rather small, however, starting at fields below 40 kOe de
Haas-van Alphen (dHvA) like oscillations in magnetostriction (Fig. \ref{F8}) are clearly observed. A fast Fourier
transform performed on these data in the form of $\Delta L/L_0$ vs. $1/H$. Seven dHvA frequencies ranging from
$\sim 1$ MG to $\sim 50$ MG were observed (Fig. \ref{F8}, inset). The occurrence of quantum oscillations in
magnetostriction is a known phenomenon \cite{cha71a}, however observations of such oscillations are rather rare,
since both large, high quality single crystals and sensitive dilatometers are required. dHvA oscillations were
observed in LuAl$_3$ via magnetostriction up to temperatures as high as 20 K (Fig. \ref{F9}). The temperature
dependence of the amplitude of these oscillations can be used to evaluate the effective masses of the
quasiparticles on the corresponding extremal orbits using standard Lifshitz - Kosevich formula
\cite{lif54a,lif55a,sho84a}. The obtained dHvA frequencies and corresponding effective masses, $m^*/m_0$, together
with the literature data \cite{sak01a} obtained by a conventional magnetic susceptibility field-modulation
technique are shown in Fig. \ref{F10}. Of the seven orbits observed in this work, four (above 10 MG) are
consistent with experimental or theoretical literature data and three frequencies ($ F < 10$ MG, $m^*/m_0 < 0.5$)
are new.
\\

\begin{figure}[tbp]
\begin{center}
\includegraphics[angle=0,width=120mm]{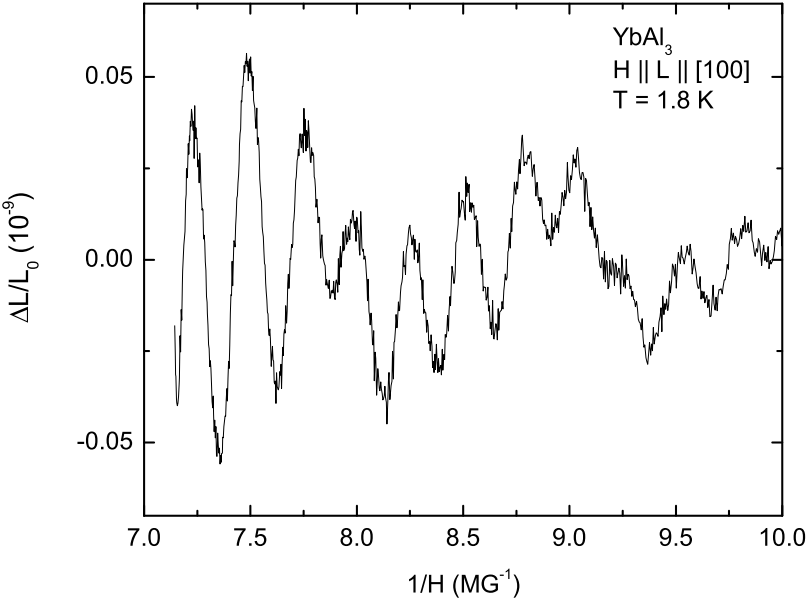}
\end{center}
\caption{Longitudinal magnetostriction in YbAl$_3$ for $H \| [100]$ at $T = 1.8$ K plotted as a function of $1/H$.
Constant background was subtracted from the data.} \label{F11}
\end{figure}

\begin{figure}[tbp]
\begin{center}
\includegraphics[angle=0,width=120mm]{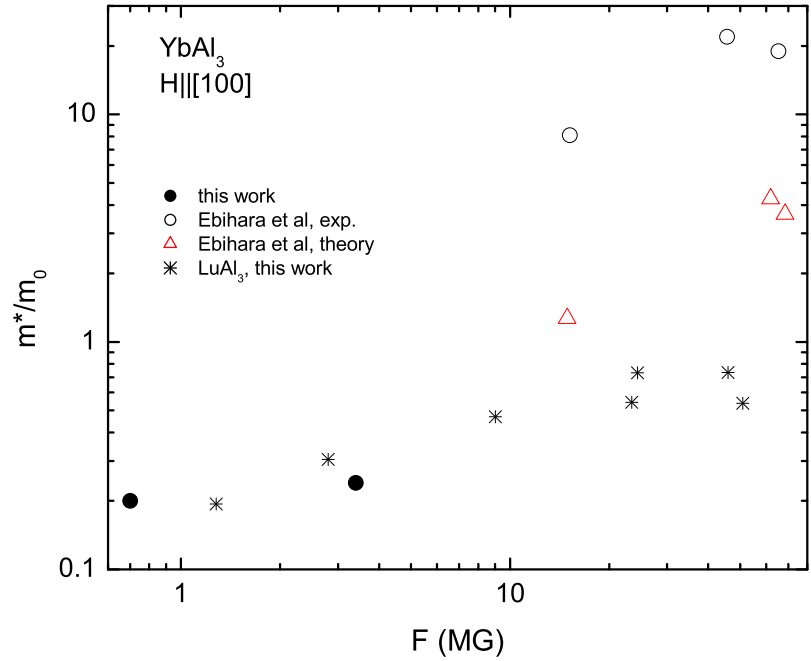}
\end{center}
\caption{(Color online) dHvA frequencies and corresponding effective masses for YbAl$_3$, $H \| [100]$. Filled
circles - this work, other symbols - literature data: open circles - experiment \cite{ebi00a}, triangles - theory
\cite{ebi00b}. Data for LuAl$_3$ (this work) are shown as asterisks for comparison.} \label{F12}
\end{figure}

The magnetostriction of YbAl$_3$ is also rather small: $\Delta L/L_0 \approx 3 \cdot 10^{-9}$ at $T = 1.8$ K, $H = 140$
kOe. In high fields  dHvA-like oscillations are also observed (Fig. \ref{F11}). The amplitude of the oscillations
is significantly smaller (few orders of magnitude for the leading frequency) than that for LuAl$_3$ and they
reduce to the level of noise above $\sim 15$ K. Two frequencies and their effective masses were identified from
our measurements and the results are plotted together with the literature data in Fig. \ref{F12}. It is worth
noting that, for the orbits detected by magnetostriction the electronic masses in YbAl$_3$ are similar to those
found for LuAl$_3$. Significant mass enhancement appears to occur only for higher frequencies.
\\

\begin{figure}[tbp]
\begin{center}
\includegraphics[angle=0,width=120mm]{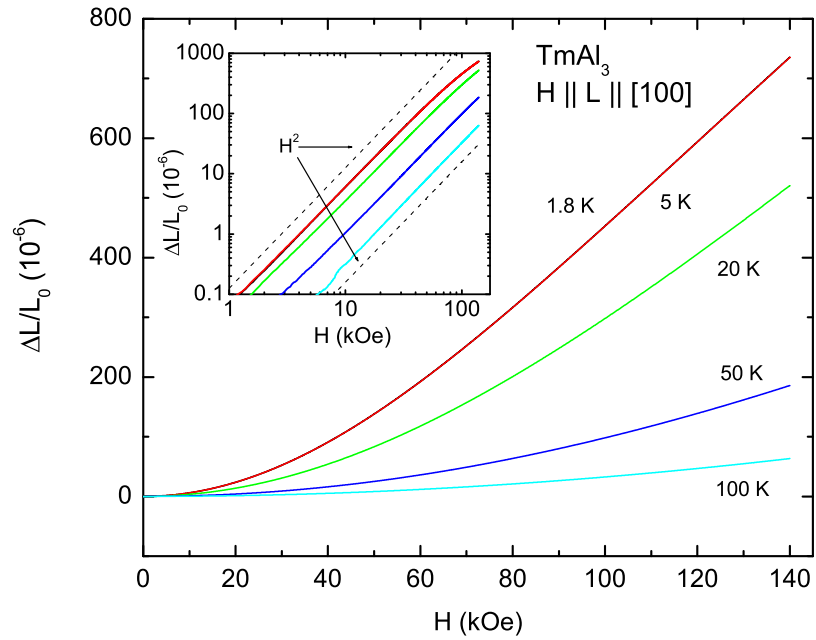}
\end{center}
\caption{(Color online) Longitudinal magnetostriction in TmAl$_3$ for $H \| [100]$ at different temperatures, from
1.8 K to 100 K. Note that within the scale of the plot, 1.8 K and 5 K data coincide. Inset: the same data on a
{\it log - log} plot. Dashed lines correspond to $\Delta L/L_0 \propto H^2$ functional behavior.} \label{F13}
\end{figure}

Longitudinal magnetostriction of TmAl$_3$ is orders of magnitude higher than that of YbAl$_3$ and LuAl$_3$:
$\Delta L/L_0 \approx 7 \cdot 10^{-4}$ at $T = 1.8$ K, $H = 140$ kOe (Fig. \ref{F13}). In the intermediate and high
fields the magnetostriction is approximately proportional to $H^2$ (Fig. \ref{F13}, inset). Detailed analysis of
the magnetostriction \cite{nic78a,cam79a,cre81a} requires knowledge of the details of the Tm$3^+$ ion-lattice
interactions.

We can not resolve quantum oscillations of magnetostriction in TmAl$_3$, although there are no indications of
inferior crystal quality. We need to mention though that the much higher, monotonic magnetostriction background
makes observation of quantum oscillations extremely difficult. Using a traditional magnetic susceptibility
modulation technique seven fundamental dHvA frequencies were observed in TmAl$_3$ for $H \| [100]$ \cite{ebi01a},
these frequencies being similar to the ones found for LuAl$_3$ and YbAl$_3$.

\section{Summary}

Temperature-dependent thermal expansion coefficients were measured between 1.87 K and 300 K for RAl$_3$ (R = Lu,
Yb, Tm) along $[100]$ direction. In YbAl$_3$ the intermediate valence of the Yb-ions results in $\alpha (T)$ for this
material being consistently lower than $\alpha (T)$ of the non-magnetic analogue, LuAl$_3$, with a negative
thermal expansion region at low temperatures. Thermal expansion coefficient, $\alpha (T)$, of TmAl$_3$ manifests
contribution from the CEF effects and, as opposed to YbAl$_3$ and LuAl$_3$, displays strong field  dependence
below $\sim 100$ K.

Magnetostriction of YbAl$_3$ and LuAl$_3$ ($H \| L \| [100]$) shows clear dHvA-like oscillations up to
temperatures as high as above 20 K for LuAl$_3$. Several new dHvA frequencies were measured (three for LuAl$_3$
and two for YbAl$_3$) and their effective masses were estimated. For these orbits the electronic masses in
YbAl$_3$ are similar to those found for LuAl$_3$.  TmAl$_3$ shows very large, exceeding that of YbAl$_3$ and
LuAl$_3$ by several orders of magnitude, magnetostriction at low temperatures.

In addition, a CEF splitting scheme for TmAl$_3$ with the $\Gamma_2$ level as the ground state is strongly
suggested based on the analysis of the heat capacity to significantly higher fields and temperatures. This result
removes a two decade ambiguity as to the CEF level scheme for this  material.

\ack

Work at Ames Laboratory was supported by the US Department of Energy - Basic Energy Sciences under Contract No.
DE-AC02-07CH11358. GMS was supported by the National Science Foundation under grant DMR-0704406. GMS also gratefully acknowledges the use of the S.I.M.W.A.P. analysis protocol.

\clearpage

\clearpage

\clearpage

\clearpage

\clearpage

\clearpage

\clearpage

\clearpage

\clearpage

\clearpage

\clearpage

\clearpage

\clearpage

\clearpage

\clearpage

\clearpage

\clearpage

\end{document}